\documentclass[fleqn, 12pt] {article}
\usepackage{epsf}
\begin{document}
\title{Effect of viscosity and surface tension on the growth of Rayleigh -Taylor instability and  Richtmyer-Meshkov instability induced two fluid interfacial nonlinear structure}
\author{M. R. Gupta\thanks{e-mail: mrgupta$_{-}$cps@yahoo.co.in}, Rahul Banerjee\thanks{e-mail:
rbanerjee.math@gmail.com},L. K. Mandal,R. Bhar,H. C. Pant,
Manoranjan Khan\thanks{e-mail: mkhan$_{-}$ju@yahoo.com} \\
Deptt. of Instrumentation Science \& Centre for Plasma Studies
\\Jadavpur University, Kolkata-700032, India\\
M. K. Srivastava\\
Theoretical Physics Division, BARC, Mumbai-400085, India\\}

\date{}

\maketitle
\begin{abstract}
The effect of viscous drag and surface tension on the nonlinear
two fluid interfacial structures induced by Rayleigh -Taylor
instability and Richtmyer-Meshkov instability are
investigated.Viscosity and surface tension play important roles on
the fluid instabilities. It is seen that the magnitude of the
suppression of the terminal growth rate of the tip of the bubble
height depends only on the viscous coefficient of the upper
(denser) fluid through which the bubble rises and surface tension
of the interface. However, in regard to spike it is shown that in
an inviscid fluid spike does not remain terminal but approaches
like a free fall under gravity as the Atwood number $(A)$
increases. In this respect there exits qualitative agreement of
our results with simulation result as also with some earlier
theoretical results. Viscosity reduces the free fall velocity
appreciably and it becomes terminal with increasing viscosity.
Results obtained from numerical integration of the relevant
nonlinear equations describing the temporal development of the
spike support the foregoing observations.
\end{abstract}
\newpage

\section*{I INTRODUCTION}
There are many causes for instability of the interface between two
fluids. Under the gravitational force when a denser fluid overlies
a lighter fluid, the instability occurs and it is called
Rayleigh-Taylor instability(RTI). Ritchmeyer-Meshkov
instability(RMI) is another type of  instability which occurs
whenever a shock front crosses the interface of two materials of
different shock impendence  (the shock must enter from the low
independence interface side). Both these instabilities play
important role in the ablation region at compression front during
the process of inertial confinement fusion, supernova remnant
formation or shock tube experiments in the laboratory[1]. In the
nonlinear regime the fluid interface forms a finger shape
structure.The structure is called a bubble (spike) if the lighter
(denser) fluid penetrates into the denser (lighter) fluid. Under
astrophysical conditions such structures may cover enormous range
of spatial distribution. Examples are suggested to be provided by
pillars ("elephant trunk") of Eagle Nebula which is identified
with the spike of a heavy fluid penetrating a lighter fluid[2-5].
Also sudden increase in the height of the ionospheric $F_{2}$
layer is caused by RTI mechanism as suggested by some
observational data[6]. Layzer[7] was first to describe the bubble
formation in a potential flow model. This model which is based on
an approximate description of the flow near the bubble tip
describes its nonlinear growth[8-13,26-28].

The effect of viscosity on Rayleigh-Taylor instability and
Richtmyer-Meshkov instability shows significant importance for
increasing wave number $k$ as $\nu k^2$ where $\nu$ is the
kinematic coefficient of viscosity.This effect is further enhanced
as $\nu$ increase with the temperature for a plasma[14]. The
importance of this feature has been discussed for supernova
remnant in Ref.[15]. In the domain of linear theory, the effect of
viscosity and surface tension on RTI was described in depth by
Chandrasekhar [16]. The same effect was described by
Mikaelian[17-18] for RMI and for RTI in finite thickness[19].
Higher magnitude of the growth rate suppression due to viscosity
was obtained in the weakly nonlinear theoretical study of Carles
and Popinet[20]. Later, a time dependent expression of the reduced
linear theory growth rate of RTI and RMI due to the combined
effect of viscosity and binary mass diffusivity between the fluids
was arrived at in linear theory by Robey[21]. The effect of
surface tension under the weakly nonlinear theory was analyzed by
Garnier et al.[22] and later Roy et al.[29] studied the same
effect in nonlinear theory.

The present paper reports the combined effect of viscosity and
surface tension on the two fluid nonlinear interfacial finger like
structures resulting due to RTI and RMI. We have analyzed the
problem based on Layzer's approach. It is seen that, in absence of
surface tension the lowering of the asymptotic velocity of the tip
of the bubble which is formed when the lighter (the lower) fluid
penetrates into the denser (the upper) fluid and thus encounters
the viscous drag due to the denser fluid,depends only on the
viscosity coefficient of denser fluid. However, in presence of
surface tension, the asymptotic velocity of the tip of the bubble
and nonlinear perturbed surface are oscillating under certain
conditions. It has been shown that, for RTI this oscillation
depends only on the surface tension but for RMI it depends on
surface tension as well as viscosity.

On the other hand it is shown that in an inviscid fluid the spikes
do not remain terminal as obtained in theoretical works[10,13]. It
is only discussed in ref.9 that the RTI spike was shown to have a
free fall $\sim\frac{1}{2}gt^2$ where $g$ is acceleration due to
gravity and the RMI spike to have a constant velocity of fall but
that too only for Atwood number $A=1$. In the present analysis the
free fall behavior is seen to hold for all values of Atwood number
$A$. The depth of the tip of the spike below the unperturbed
interfacial surface is $\sim\frac{1}{2}\gamma gt^2$ where $\gamma$
is a dimensionless constant which $\rightarrow1$ as
$A\rightarrow1$. Similar result is found to hold for RMI. The
effect of viscosity is seen to reduce the spike velocity
appreciably and as the viscosity coefficient increases the spike
velocity tends to become approximately terminal. Such behavior of
both RTI and RMI spike for for inviscid and viscous fluids was not
found earlier.

  The paper is organized as follows: In Section II, we describe the basic fluid equation
with the assumption that the motion is irrotational and fluid is
incompressible. Also the kinematic and dynamical boundary
condition are derived in Sec.II. The analytical expressions of the
viscosity and surface tension induced asymptotic velocity of the
tip of the bubble and associated numerical results are given in
Section III. Section IV is devoted to temporal development of the
spike. The results are summarized in Section V.

\section*{II BASIC EQUATION AND BOUNDARY CONDITIONS }

The $x-y$ plane ($z=0$) is assumed to be the unperturbed interface
between the denser fluid of density $\rho_h$ (region $z>0$) and
lighter fluid of density $\rho_l$ (region $z<0$). The variables
with subscript $h$ and $l$ represents denser and lighter fluid,
respectively. Gravity $g$ is taken to point along negative
$z$-axis. After perturbation the finger shape interface is assumed
to take up a parabolic form, given by
\begin{eqnarray} \label{eq:1}
z=\eta(x,t)=\eta_0(t)+\eta_2(t)x^2
\end{eqnarray}
\begin{eqnarray}\label{eq:2}
\mbox{For a bubble:}\qquad\qquad \eta_0>0 \quad \mbox{and} \quad
\eta_2<0,
\end{eqnarray}
\begin{eqnarray}\label{eq:3}
\mbox{For a spike:} \qquad\qquad \eta_0<0 \quad \mbox{and} \quad
\eta_2>0.
\end{eqnarray}
Following Goncharov [13], the velocity potentials describing the
irrotational motion for the denser and lighter fluids are assumed
to be given by
\begin{eqnarray}\label{eq:4}
\phi_h(x,z,t)=a_1(t)\cos{(kx)}e^{-k(z-\eta_0(t))}; \quad z>0
\end{eqnarray}
\begin{eqnarray}\label{eq:5}
\phi_l(x,z,t)=b_0(t)z+b_1(t)\cos{(kx)}e^{k(z-\eta_0(t))}; \quad
z<0
\end{eqnarray}
where $k$ is the perturbed wave number.

To find the five unknown functions i.e.
$\eta_0(t)$,$\eta_1(t)$,$a_1(t)$,$b_0(t)$ and $b_1(t)$, we require
as many equations obtained from the kinematical and dynamical
boundary conditions describing the dynamics.

We first turn to the kinematical boundary conditions corresponding
to the interfacial surface perturbations represented by eq.(1):
\begin{eqnarray}\label{eq:6}
\eta_x(v_h)_{x}-\eta_x(v_l)_{x}=(v_h)_{z}-(v_l)_{z}
\end{eqnarray}
\begin{eqnarray}\label{eq:7}
\eta_t+\eta_x(v_h)_{x}=(v_h)_{z}
\end{eqnarray}
Substituting eq.(6) and eq.(7)from eq.(1) and for
$(v_{h(l)})_x=-\frac{\partial \phi_{h(l)}}{\partial x}$ and
$(v_{h(l)})_x=-\frac{\partial \phi_{h(l)}}{\partial x}$ from
eq.(4)and eq.(5) and expanding in powers of the transverse
coordinate $x$ neglecting terms O($x^{i}$)($i\geq3$), we obtain
the following relations which are equivalent to the kinematic
boundary conditions eq.(6) and eq.(7):
\begin{eqnarray}\label{eq:8}
\frac{d\xi_1}{d\ t}=\xi_3
\end{eqnarray}
\begin{eqnarray}\label{eq:9}
\frac{d\xi_2}{d\ t}=-\frac{1}{2}(6\xi_2+1)\xi_3
\end{eqnarray}
\begin{eqnarray}\label{eq:10}
b_0=-\frac{6\xi_2}{3\xi_2-\frac{1}{2}}ka_1
\end{eqnarray}
\begin{eqnarray}\label{eq:11}
b_1=\frac{3\xi_2+\frac{1}{2}}{3\xi_2-\frac{1}{2}}a_1
\end{eqnarray}
\begin{eqnarray}\label{eq:12}
\mbox{where }\xi_1=k\eta_0; \qquad \xi_2=\eta_2/k; \qquad
\xi_3=k^2a_1; \qquad \tau=t\sqrt{kg}
\end{eqnarray}
$\xi_1$ and $\xi_2$ are, respectively, the non-dimensionalized
displacement and curvature at the tip of the nonlinear structure,
$\xi_3/k$ is its velocity and $\tau$ is the non-dimensionalized
time. Eq.(8) and eq.(9) are the first two of the three time
development equations needed to describe the time evaluation of
the nonlinear structure (the other two viz $b_0(t)$ and $b_1(t)$
are provided by eq.(10) and eq.(11)) .

For the constant density fluid, the equation of continuity is
$\vec\nabla.\vec{v}=0$, which becomes $\nabla^{2}\phi=0$ for
potential flow. So, for a fluid with uniform viscosity having
coefficient of viscosity $\mu_{h(l)}$, the viscous term drops out
($\mu\nabla^{2}\vec{v}=\mu\vec\nabla(\nabla^{2}\phi)=0$) in
\begin{eqnarray}\label{eq:13}
\rho\left[\frac{\partial \vec{v}}{\partial
t}+(\vec{v}.\vec{\nabla})\vec{v}\right]=-\vec{\nabla}p+\mu\nabla^{2}\vec{v}
-\rho g\hat{z}
\end{eqnarray}
and we arrive at the first integral of the momentum equation.
\begin{eqnarray}\label{eq:14}
-\rho_{h(l)}\frac{\partial \phi_{h(l)}}{\partial t}+
\frac{1}{2}\rho_{h(l)}(\vec{\nabla} \phi_{h(l)})^{2}+\rho_{h(l)} g
z=-p_{h(l)}+f_{h(l)}(t)
\end{eqnarray}
The net stress [16] at two fluid interface including that due to
viscosity is
\begin{eqnarray}\label{eq:15}
P_{h(l)}=-p_{h(l)}+2\mu_{h(l)}\frac{\partial
(v_{h(l)})_z}{\partial t}
\end{eqnarray}
Plugging the dynamical boundary condition $P_h-P_l=-T/R$ at the
interface $z(x,t)=\eta(x,t)$, where $R $ is the radius of
curvature and $T$ is the surface tension of the perturbed
interface, in eq.(14) and eq.(15) we obtained the following
equation[23].
\begin{eqnarray}\label{eq:16}
\rho_h[-\frac{\partial\phi_h}{\partial t}+\frac{1}{2}
(\vec{\nabla}\phi_h)^2]-\rho_l [-\frac{\partial \phi_l}{\partial
t}+\frac{1}{2}(\vec{\nabla}
\phi_l)^2]+g(\rho_h-\rho_l)z=2[\mu_h\frac{\partial^2
\phi_h}{\partial z^2}-\mu_l\frac{\partial^2 \phi_l}{\partial
z^2}]-\frac{T}{R}+ f_h-f_l
\end{eqnarray}
at the interface $z(x,t)=\eta(x,t)$.

Substituting for $\phi_h$,$\phi_l$ from eq.(4),eq.(5) and value of
$\frac{1}{R}$, using eq.(10)-eq.(12) and equating coefficient of
$x^2$, we obtain after some straightforward but lengthy algebraic
manipulation, the following time development equation for $\xi_3$:
\begin{eqnarray}\label{eq:17}
\nonumber\frac{d(\xi_3/\sqrt{kg})}{d\tau}=-\frac{N(\xi_2,r)}{D(\xi_2,r)}\frac{(\xi_3/\sqrt{kg})^2}{(6\xi_2-1)}+2(r-1)\frac{\xi_2(6\xi_2-1)}{D(\xi_2,r)}(1-12\xi_2^2\frac{k^2}{k_c^2})\\
-\frac{2(\xi_3/\sqrt{kg})}{D(\xi_2,r)}r
c_h[(s+1)(1-12\xi_2^2)+4\xi_2(s-1)]\hskip 50pt
\end{eqnarray}
where
\begin{eqnarray}\label{eq:18}
\quad
r=\frac{\rho_h}{\rho_l};\nu_{h(l)}=\frac{\mu_{h(l)}}{\rho_{h(l)}};s=\frac{\mu_l}{\mu_h};c_h=\frac{k^{2}\nu_h}{\sqrt{kg}};k_c^2=\frac{(\rho_h-\rho_l)g}{T}
\end{eqnarray}
\begin{eqnarray}\label{eq:19}
\nonumber D(\xi_2,r)=12(1-r)\xi_{2}^{2}+4(1-r)\xi_{2}+(r+1); \\
N(\xi_2,r)=36(1-r)\xi_{2}^{2}+12(4+r)\xi_{2}+(7-r)
\end{eqnarray}
Eq.(8) and eq.(9) together with eq.(17) governs the temporal
development of the Rayleigh-Taylor instability. For
Richtmyer-Meshkov instability, the gravity dependent term in
eq.(17) vanishes i.e. $g=0$ and the equation for
$\frac{d\xi_3}{dt}$ becomes
\begin{eqnarray}\label{eq:20}
\nonumber
\frac{d(\xi_3/\sqrt{kg})}{d\tau}=-\frac{N(\xi_2,r)}{D(\xi_2,r)}\frac{(\xi_3/\sqrt{kg})^2}{(6\xi_2-1)}-\frac{24(r-1)\xi_2^3(6\xi_2-1)}{D(\xi_2,r)}\frac{k^2}{k_c^2}\\
 -\frac{2(\xi_3/\sqrt{kg})}{D(\xi_2,r)}r c_h
[(s+1)(1-12\xi_2^2)+4\xi_2(s-1)] \hskip 1pt
\end{eqnarray}

\section*{III SUPPRESSION AND OSCILLATION OF ASYMPTOTIC BUBBLE VELOCITY }

The nondimensionalized time development plots of $\xi_1$,$\xi_2$
and $\xi_3$ for RTI bubble are shown in Figure 1. As
$\tau\rightarrow\infty$, the asymptotic values of $\xi_2$ and
$\xi_3$ for bubble are obtained by setting
$\frac{d\xi_2}{d\tau}=0$ giving
$[(\xi_2)_{asymp}]_{bubble}=-\frac{1}{6}$ and
$\frac{d(\frac{\xi_3}{\sqrt{kg}})}{d\tau}=0$ yielding
\begin{eqnarray}\label{eq:21}
[(\xi_3)_{asymp}]_{bubble}=\frac{\frac{2}{3}
\frac{A}{1+A}kg(1-\frac{k^2}{3k_c^2})}{\sqrt{\frac{4}{9}{\nu_h}^2k^4
+ \frac{2}{3} \frac{A}{(1+A)}kg(1-\frac{k^2}{3k_c^2})}+
\frac{2}{3}\nu_hk^2}
\end{eqnarray}

It is interesting to note that if $k^2<3k_c^3$, the asymptotic
velocity of the bubble caused by the rising of the lighter (lower)
fluid also by pushing through the denser (upper) fluid is affected
only by the viscous drag exerted by the later.This is clearly seen
from eq.(21) as $[(\xi_3)_{asymp}]_{bubble}$ depends only on the
kinematic viscosity of the denser (upper) fluid ($\nu_h$,). If
$k^2=3k_c^2$ equilibrium is attained; but if $k^2>3k_c^2$, this
reverse the sign of the second term in eq.(17) leads to the
emergence of oscillatory state (Figure 2).

Similar results for temporal development of Richtmyer-Meshkov
instability are shown in Figure 3. The asymptotic velocity is
\begin{eqnarray}\label{eq:22}
[(\xi_3)_{asymp}]_{bubble}=-\frac{2k^2\nu_h}{3}+\sqrt{\frac{4}{9}\nu_h^2k^4-kg\frac{2A}{9(1+A)}\frac{k^2}{k_c^2}}\coth{[(\frac{3(1+A)}{3+A}\sqrt{\frac{4}{9}\nu_h^2k^4-kg\frac{2A}{9(1+A)}\frac{k^2}{k_c^2}})t]}
\end{eqnarray}

For bubble as well as for spike,
$[(\xi_3)_{asymp}]_{bubble}\rightarrow-\frac{2k^2\nu_{h}}{3}+\sqrt{\frac{4}{9}\nu_{h}^2k^4-kg\frac{2A}{9(1+A)}\frac{k^2}{k_c^2}}$
exponentially with time. In absence of surface tension the time
dependence is qualitatively similar to that for linear theoretical
result of Mikaelian[17] but with different Atwoood number and
kinematic viscosity coefficient dependence. However, the weakly
nonlinear theoretical results of Carles and Popinet[20] shows a
different time dependence $~a[1-\frac{4}{3\sqrt{\pi}}k\sqrt{\nu
t}]$. On the other hand in case of RMI,for $ k^2>2c_{h(l)}^2
k_c^2\frac{1+A}{A}$ the asymptotic velocity of the bubble as well
as the perturbed surface elevation oscillates. This is represented
in Figure 4.

In absence of surface tension both RTI and RMI are characterized
by linear inviscid growth rate which increase with increasing wave
number $k$. The dissipative effect due to viscosity also increases
with increasing $k$ and suppress the growth rate. A graphical
representation of the wave number dependence of the nonlinear
growth rate of the hight of the tip of the bubble is shown in
Figure 5. The nature of $k-$ dependence is qualitatively similar
to that in linear case [21] except that for $t\rightarrow \infty$
where the growth rate tends to a saturation value for all $k$ in
the nonlinear case.For bubble the asymptotic growth rate for RTI
is maximum at
$[k_{max}]_{bubble}=\sqrt[3]{\frac{3A}{16(1+A)}\frac{g}{\nu_h^2}}$.

Thus for RTI the growth rate and perturbed interface are
oscillating due to surface tension while for RMI oscillation
depends on the relative strength of surface tension and viscous
drag.

\section*{IV TIME DEVELOPMENT OF SPIKE}

To study the time evolution of spikes we adopt a procedure
different from the usual Goncharov transformation[13]. We cast
time evolution eq.(17) in the form given bellow.
\begin{eqnarray}\nonumber
\frac{d(\xi_3/\sqrt{kg})}{d\tau}=-\frac{1}{2(\xi_2-\frac{1}{6})(\xi_2-\beta_{+})(\xi_2-\beta_{-})}
[(\xi_2-\alpha_{+})(\xi_2-\alpha_{-})(\xi_3/\sqrt{kg})^2+2\xi_2(\xi_2-\frac{1}{6})^{2}
(1-12\xi_2^2\frac{k^2}{k_c^2})
\end{eqnarray}
\begin{eqnarray}\label{eq:23}
\hskip 73pt -\frac{1}{3}(\xi_2-\frac{1}{6})r
c_{h}[(s+1)(1-12\xi_2^2)+4\xi_2(s-1)](\xi_3/\sqrt{kg})]
\end{eqnarray}
by using the following expression for $N(\xi_{2},r)$ and
$D(\xi_{2},r)$:
\begin{eqnarray}\label{eq:24}
\nonumber
N(\xi_{2},r)=-36(r-1)(\xi_2-\alpha_{+})(\xi_2-\alpha_{-})\\
D(\xi_{2},r)=-12(r-1)(\xi_2-\beta_{+})(\xi_2-\beta_{-})
\end{eqnarray}
\begin{eqnarray}\label{eq:25}
\mbox{where }\nonumber \alpha_{\pm}=\frac{(r+4)\pm \sqrt{16r+4}}{6(r-1)}\\
\beta_{\pm}=\frac{-1\pm \sqrt{\frac{4r+2}{r-1}}}{6}\hskip 35pt
\end{eqnarray}
Clearly $\alpha_{+}>\alpha_{-}$ and $\beta_{+}>\beta_{-}$ and also
$\alpha_{+}>\beta_{+}$, $\alpha_{+}>\frac{1}{6}$

First we consider the temporal development of spike as a result of
RTI in an inviscid fluid with absence of surface tension, i.e., we
put $c_{h}=0$ and $\frac{1}{k_{c}^{2}}=0$ in eq.(17). We start
with an initial value of $\xi_{2}=\xi_{20}>\alpha_{+}>\beta_{+}$
and $\frac{1}{6}$, and $\xi_{30}<0$.

Eq.(9) shows that $\frac{d\xi_{2}}{d\tau}>0$ while eq.(23) shows
that $\frac{d(\xi_{3}/\sqrt{kg})}{d\tau}<0$ (as the square
bracketed term on the RHS of the latter equation is positive) for
all $\tau$ when one starts from such initial values. Thus
$\xi_{2}(\tau)$ is a monotonically increasing and
$\xi_{3}/\sqrt{kg}$ is a monotonically decreasing function of time
$\tau$. Now the curvature at the tip of the spike, i.e., $x=0$ is
$\frac{1}{R}=\frac{\partial^{2}\eta}{\partial
x^{2}}/[1+(\frac{\partial \eta}{\partial
x})^{2}]^{\frac{3}{2}}=2\eta_{2}(\tau)=k\xi_{2}(\tau)$. Thus the
curvature of the spike increases with time while the acceleration
of the tip of spike $\frac{d(\xi_{3}/\sqrt{kg})}{d\tau}$ (directed
downward) tends to a constant value as may be interred from
eq.(23) when viscosity and surface tension are neglected. Thus the
spike appears to fall continuously and simultaneously gets
sharpened. This result is quite different from the earlier results
obtained by Goncharov's transformation which concludes that the
spike velocity tends asymptotically to a constant value. Our
result is in conformity with expected spike behavior and is in
agreement with some earlier theoretical results obtained by a
different approximate method[24][25]. The same qualitative spike
behavior is exhibited in presence of viscosity but with much
reduced speed of fall of the tip of the spike. This is
demonstrated in Figure 6 and Figure 7 obtained from the results
derived from numerical integration of eq.(8), eq.(9) and eq.(23).
Figure 6 shows that the spike speed decreases as the coefficient
of viscosity increases. Moreover, for $c_{h}=0$ (inviscid fluid)
the spike velocity is seen to vary linearly with time (i.e., close
to free fall velocity) so that the displacement of the tip of the
spike $\sim\frac{1}{2} \gamma g t^{2} $ where $\gamma $ is a
dimensionless constant and value of $\gamma$ close to unity as
$A\rightarrow1$.

Rayleigh-Taylor instability is driven by gravity $g$ while
Richtmyer-Meshkov instability is switched on by the impingement of
a shock which impulsively changes the normal velocity by the
amount $\Delta v= v_{after}-v_{before}$. Thus Richtmyer-Meshkov
instability is driven by the instantaneous acceleration $\triangle
v \delta (t)$. This has the consequence that the dynamical
variables are to be nondimensionalized using normalization in
terms of $(k\triangle v)$ for RMI instead of $\sqrt{kg}$ for RTI.
Hence, in RMI equations $\frac{\xi_{3}}{\sqrt{kg}}$, $c_{h}$ and
$\tau$ are replaced by
\begin{eqnarray}\label{eq:26}
\frac{\xi_{3}}{\sqrt{kg}}=\frac{\xi_{3}}{(k\Delta v)}, \mbox{ }
\overline{c}_{h}=\frac{k^2 \nu_{h}}{(k\triangle v)},\mbox{      }
\overline{\tau}=t(k\Delta v)
\end{eqnarray}

With replacements as gravity by eq.(26), the temporal development
of RMI spike growth are obtained from numerical integration of
eq.(8), eq.(9) and eq.(23) when the gravity $g$ induced second
term in the square bracket on the RHS of the last mentioned
equation is to be deleted. The absence of gravity induced
acceleration keeps $\frac{\xi_{3}}{(k\Delta v)}$ close to its
initial value when viscosity is neglected $(\overline{c}_{h}=0)$
but tends to vanish when $\overline{c}_{h}\neq 0$.

The lowering of the value of $\frac{\xi_{3}}{(k\Delta v)}$ in
magnitude will according to eq.(9) reduce the rate of growth of
the curvature $\xi_{2}$ with respect to $\overline{\tau}$
(nondimensionalized time). Thus, in contrast to the free fall of
RTI spike, the RMI spike tip descends at an almost constant rate.
The curvature also increases slowly, i.e, the spike sharpens at a
much reduced rate as it falls (as compared to the RTI spike). All
these features are shown in figure 8 and figure 9. The growth rate
are seen to be further reduced due to the viscous drag.

Finally, we note the difference between the RTI and RMI parabolic
spike structures (figure 7 and figure 9) given by
\begin{eqnarray}\label{eq:27}
\nonumber ky=k\eta_{0}(t)+(\frac{\eta_{2}(t)}{k})(kx)^2\\
=\xi_{1}(t)+\xi_{2}(t) (kx)^2 \hskip 19 pt
\end{eqnarray}
Let us first consider the inviscid case. The discussions in the
foregoing paragraphs indicate that
\begin{eqnarray}\label{eq:28}
\xi_{1}(\tau)\sim - \frac{1}{2} \tau^2 \mbox{ for RTI while }
\xi_{1}(\overline{\tau})\sim -\mbox{(constant) }\overline{\tau}
\mbox{  for RMI }
\end{eqnarray}
Further from eq.(8) and eq.(9) which gives
$(\xi_{2}+\frac{1}{6})=(\xi_{20}+\frac{1}{6})exp
[-3(\xi_{1}-\xi_{10})]$ ($\xi_{10}$ and $\xi_{20}$ are initial
values) we obtain
\begin{eqnarray}\label{eq:29}
\xi_{2}(\tau)\sim  \mbox{(constant) } \exp [\frac{3}{2} \tau^2]
\mbox{ for RTI while } \xi_{2}(\overline{\tau})\sim \mbox{
(constant) } \exp [\mbox{ (constant) } \overline{\tau}]\mbox{  for
RMI }
\end{eqnarray}

Eq.(28) and eq.(29) when plugged in eq.(27), now clearly indicates
that the RTI spike fall much faster and gets sharpened much more
rapidly. In both cases, the effect of viscosity is to dampen the
growth rates. This leads to difference in the RTI and RMI spike
structures as seen in figure 7 and figure 9.

\section*{V SUMMARY }

Finally we briefly summarize the results:
\newline
(i) For RTI the asymptotic velocity of the tip of the bubble is
given by \\$[(\xi_3)_{asymp}]_{bubble}=$ $\frac{\frac{2}{3}
\frac{A}{1+A}kg(1-\frac{k^2}{3k_c^2})}{\sqrt{\frac{4}{9}{\nu_h}^2k^4
+ \frac{2}{3} \frac{A}{(1+A)}kg(1-\frac{k^2}{3k_c^2})}+
\frac{2}{3}\nu_hk^2} $ \\and for RMI, the asymptotic velocity of
the bubble is given by
\\$[(\xi_3)_{asymp}]_{bubble}=-\frac{2k^2\nu_h}{3}+\sqrt{\frac{4}{9}\nu_h^2k^4-kg\frac{2A}{9(1+A)}\frac{k^2}{k_c^2}}$
$\coth{[(\frac{3(1+A)}{3+A}\sqrt{\frac{4}{9}\nu_h^2k^4-kg\frac{2A}{9(1+A)}\frac{k^2}{k_c^2}})t]}$.

(ii) In case of RTI, if $k^2<3k_c^2$ the asymptotic velocity of
the bubble is affected only by the viscous drag of the upper
fluid(Figure 1) and similar effect for RMI with $k^2<2c_{h}^2
k_c^2\frac{1+ A}{A}$(Figure 3).

(iii) For RTI, the growth rate and perturbed surface are
oscillating if $k^2>3k_c^2$, i.e the oscillation depends only on
the surface tension of the perturbed interface (Figure 2) while
for RMI the growth rate and perturbed surface are oscillating if
$k^2>2c_{h}^2 k_c^2\frac{1+A}{A}$, i.e the oscillation depends on
the surface tension of the perturbed interface as well as the
coefficient of viscosity (Figure 4).

(iv) In inviscid fluid the RTI spike has no asymptotically
terminal velocity. Rather the spike has a nearly free fall so that
the depth of the tip of the spike below unperturbed surface of
separation $\sim \frac{1}{2} \gamma g t^{2} $ where $\gamma $ is a
dimensionless constant and value of $\gamma$ close to unity as
$A\rightarrow1$; also the spike sharpens as it falls. For viscous
fluid the velocity of fall gets reduced as the coefficient of
viscosity increases and tends to a nearly terminal velocity for
sufficiently large viscosity. Similar result holds for RMI. The
results are demonstrated in Figure 8 and Figure 9.

\section*{ACKNOWLEDGMENTS }
This work is supported by the C.S.I.R, Government of India under
ref. no. R-10/B/1/09.

\begin{figure}[p]
\vbox{\hskip 1.cm \epsfxsize=12cm \epsfbox{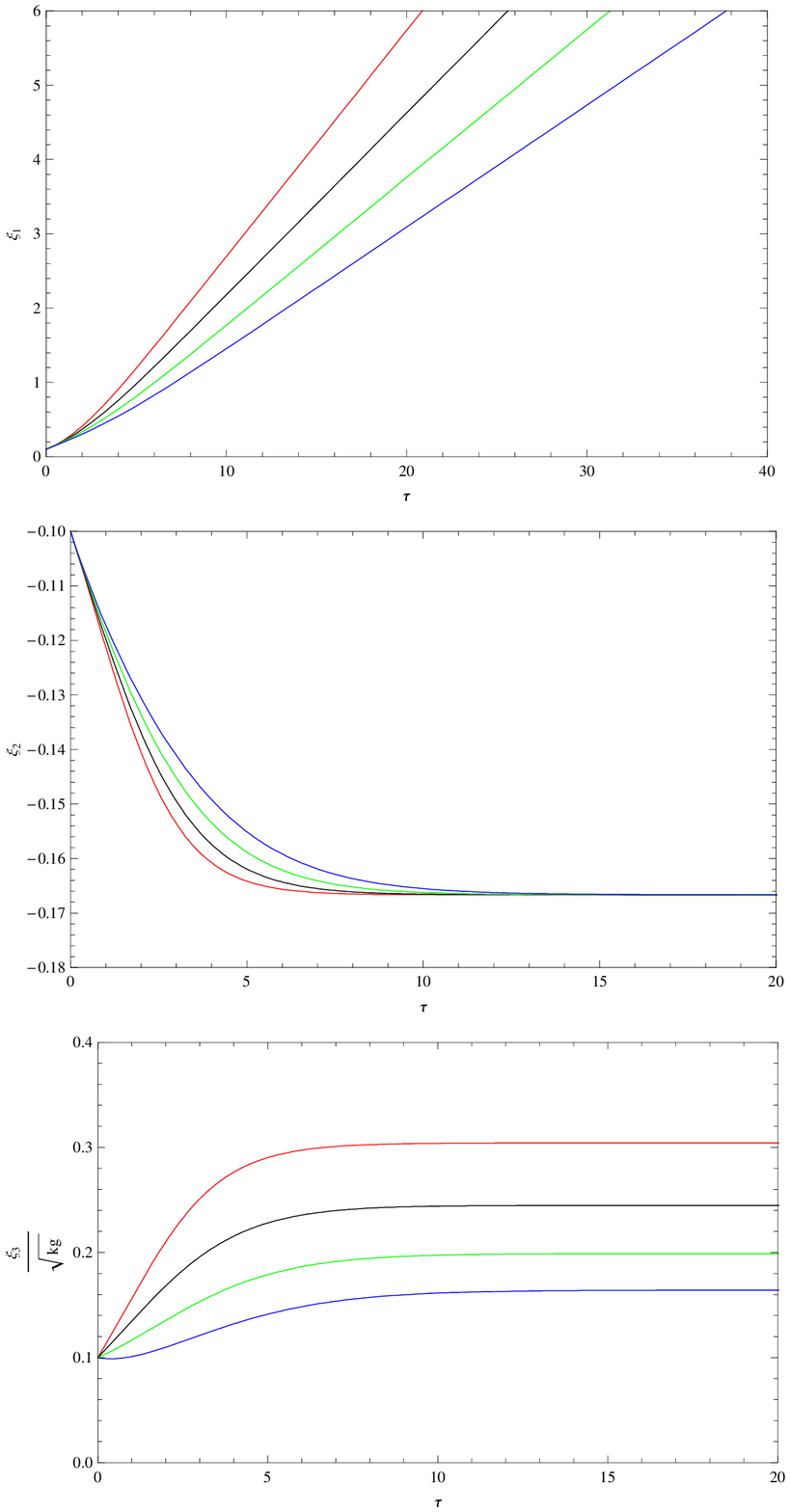}}
\begin{verse}
\vspace{-0.1cm} \caption{Variation of  $\xi_1$,$\xi_2$ and
$\xi_3/\sqrt{kg}$ with $\tau$ as obtained by the solution of the
eq.(8),eq.(9) and eq.(17) for RTI bubble with initial value
$\xi_1=0.1$, $\xi_1=-0.1$, $\xi_3/\sqrt{kg}=0.1$, $r=1.5$,
$\frac{k^2}{k_c^2}=0.5$ and $c_h$= 0 (Red), 0.1(Black),
0.2(Green), 0.3(Blue).}\label{Fig:1}
\end{verse}
\end{figure}

\begin{figure}[p]
\vbox{ \hskip 1.cm \epsfxsize=12cm \epsfbox{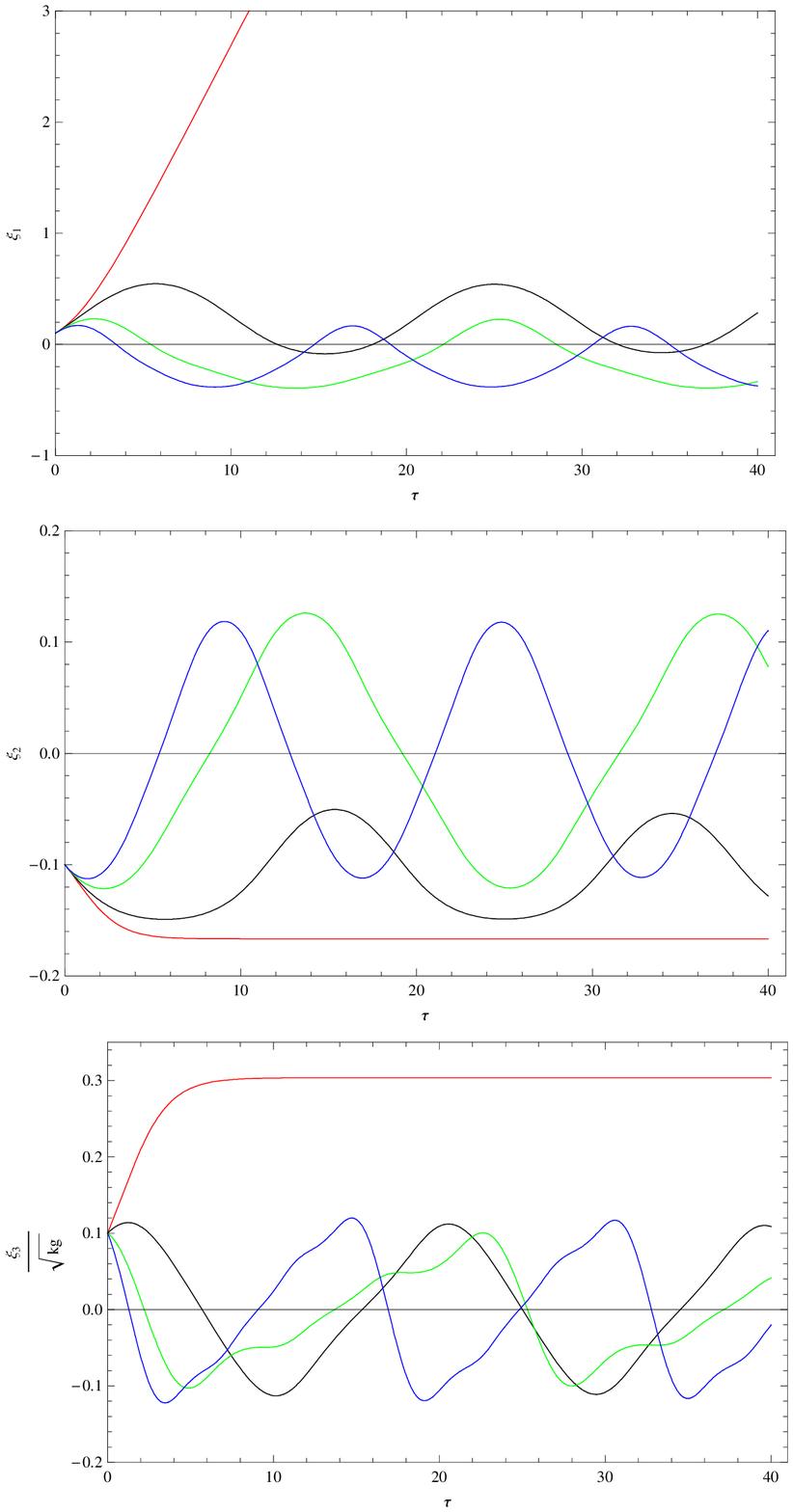}}
\begin{verse}
\vspace{-0.1cm} \caption{Variation of  $\xi_1$,$\xi_2$ and
$\xi_3/\sqrt{kg}$ with $\tau$ as obtained by the solution of the
eq.(8),eq.(9) and eq.(17) for RTI bubble with initial value
$\xi_1=0.1$, $\xi_1=-0.1$, $\xi_3/\sqrt{kg}=0.1$, $r=1.5$,
$c_h=0.001$ and $\frac{k^2}{k_c^2}$= 0.5 (Red), 5(Black),
10(Green), 15(Blue).} \label{Fig:2}
\end{verse}
\end{figure}

\begin{figure}[p]
\vbox{\hskip 1.cm \epsfxsize=12cm \epsfbox{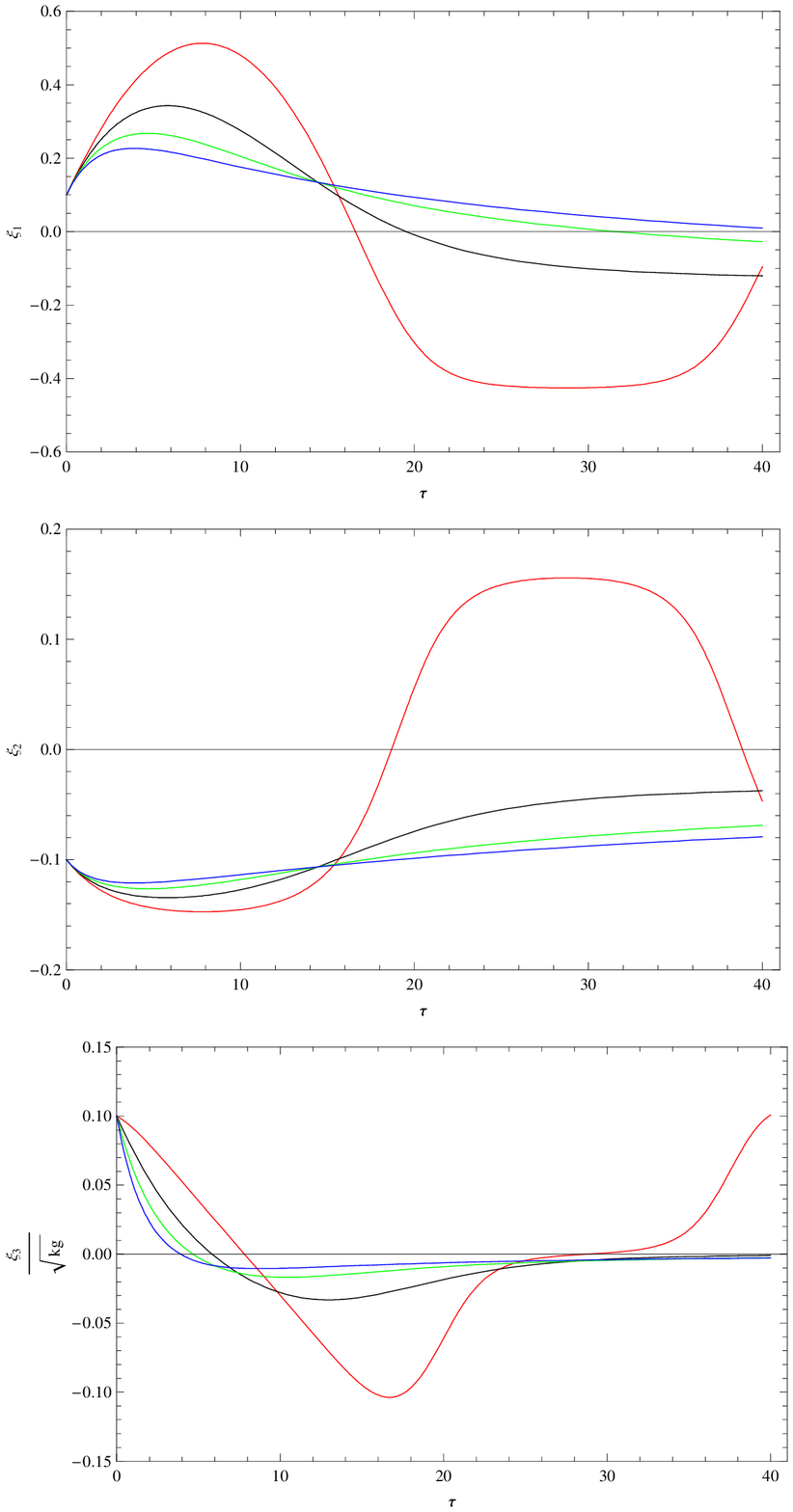}}
\begin{verse}
\vspace{-0.1cm} \caption{Variation of  $\xi_1$,$\xi_2$ and
$\xi_3/\sqrt{kg}$ with $\tau$ as obtained by the solution of the
eq.(8),eq.(9) and eq.(20) for RMI bubble with initial value
$\xi_1=0.1$, $\xi_1=-0.1$, $\xi_3/\sqrt{kg}=0.1$, $r=1.5$,
$\frac{k^2}{k_c^2}=0.5$ and $c_h$= 0 (Red), 0.1(Black),
0.2(Green), 0.3(Blue).}\label{Fig:3}
\end{verse}
\end{figure}

\begin{figure}[p]
\vbox{ \hskip 1.cm \epsfxsize=12cm \epsfbox{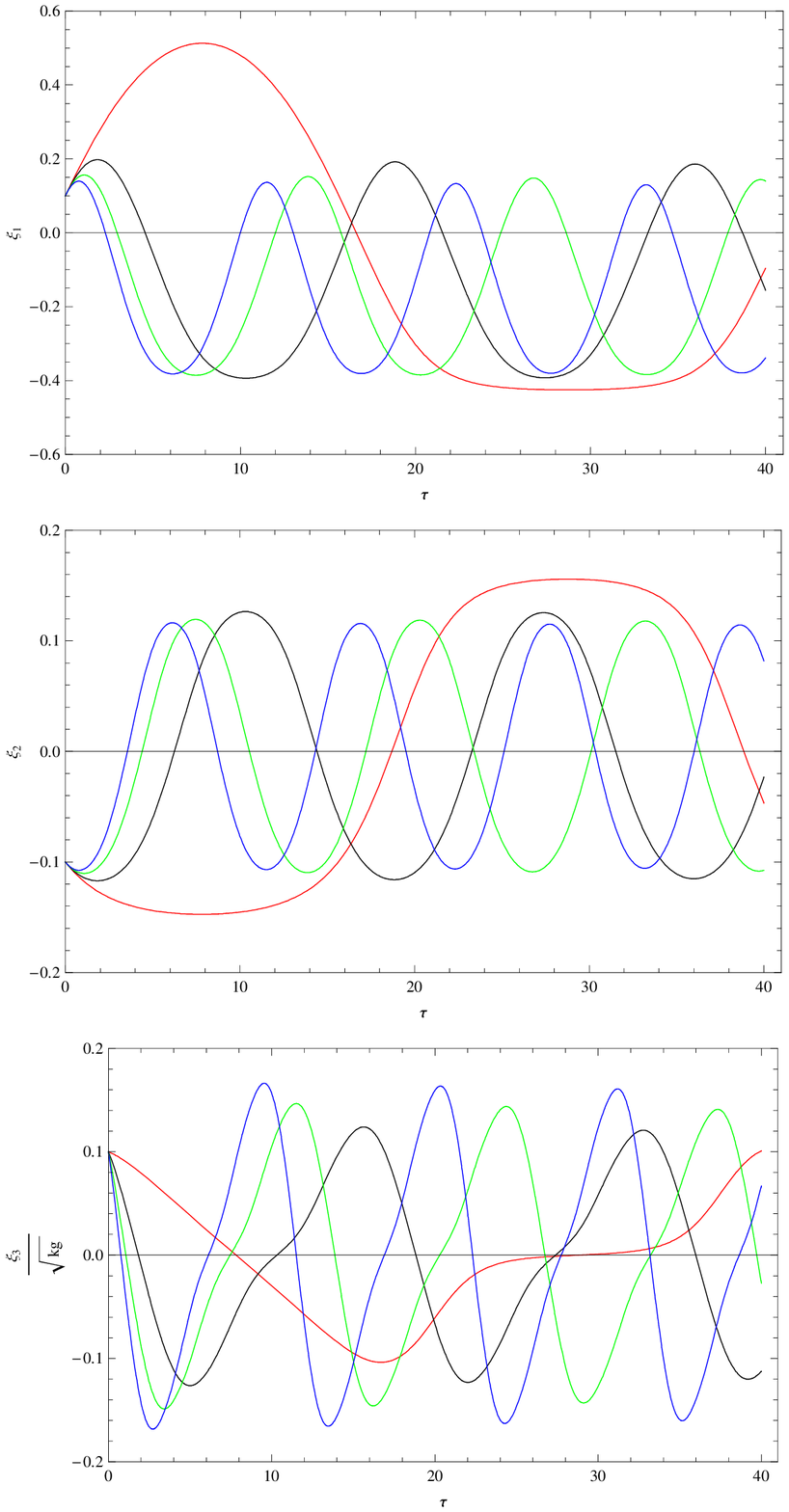}}
\begin{verse}
\vspace{-0.1cm} \caption{Variation of  $\xi_1$,$\xi_2$ and
$\xi_3/\sqrt{kg}$ with $\tau$ as obtained by the solution of the
eq.(8),eq.(9) and eq.(20) for RMI bubble with initial value
$\xi_1=0.1$, $\xi_1=-0.1$, $\xi_3/\sqrt{kg}=0.1$, $r=1.5$,
$c_h=0.001$ and $\frac{k^2}{k_c^2}$= 0.5 (Red), 5(Black),
10(Green), 15(Blue).} \label{Fig:4}
\end{verse}
\end{figure}

\begin{figure}[p]
\vbox{ \hskip 1.cm \epsfxsize=12cm \epsfbox{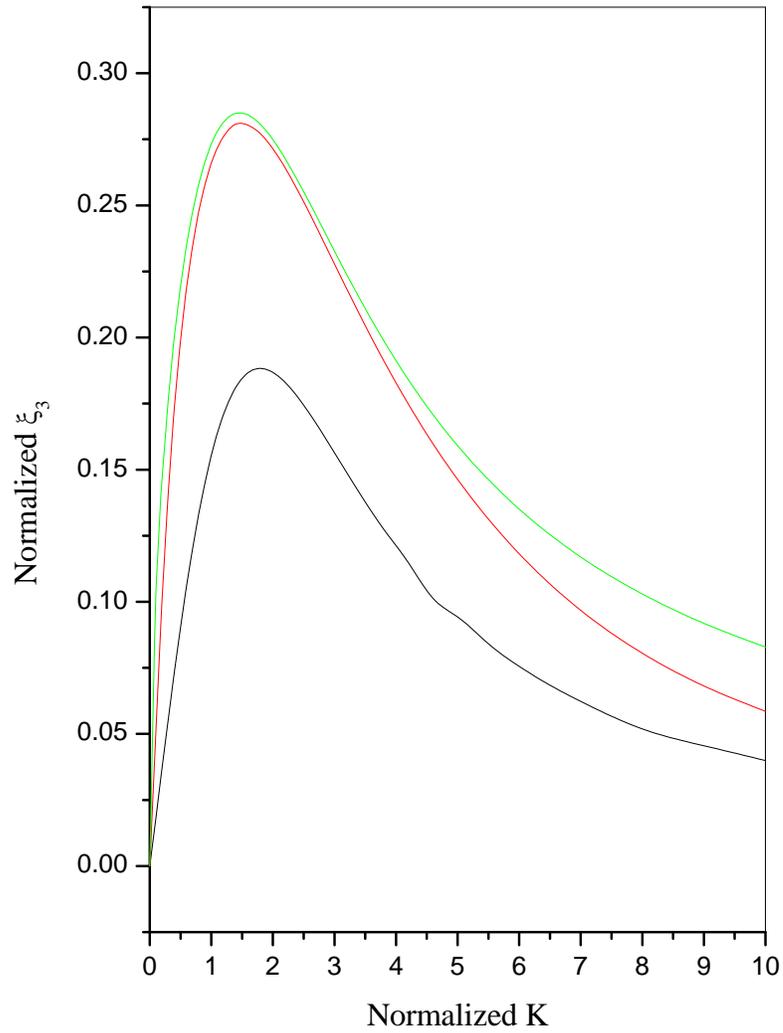}}
\begin{verse}
\vspace{-0.1cm} \caption{Variation of normalize $\xi_3$ with
normalize 'k' for RTI bubble with $c_h=0.1$, $r=1.5$,
$\frac{k^2}{k_c^2}$= 0 and $\tau=$ 3(Black), 8(Red), $\infty$
(Green).} \label{Fig:5}
\end{verse}
\end{figure}

\begin{figure}[p]
\vbox{ \hskip 1.cm \epsfxsize=12cm \epsfbox{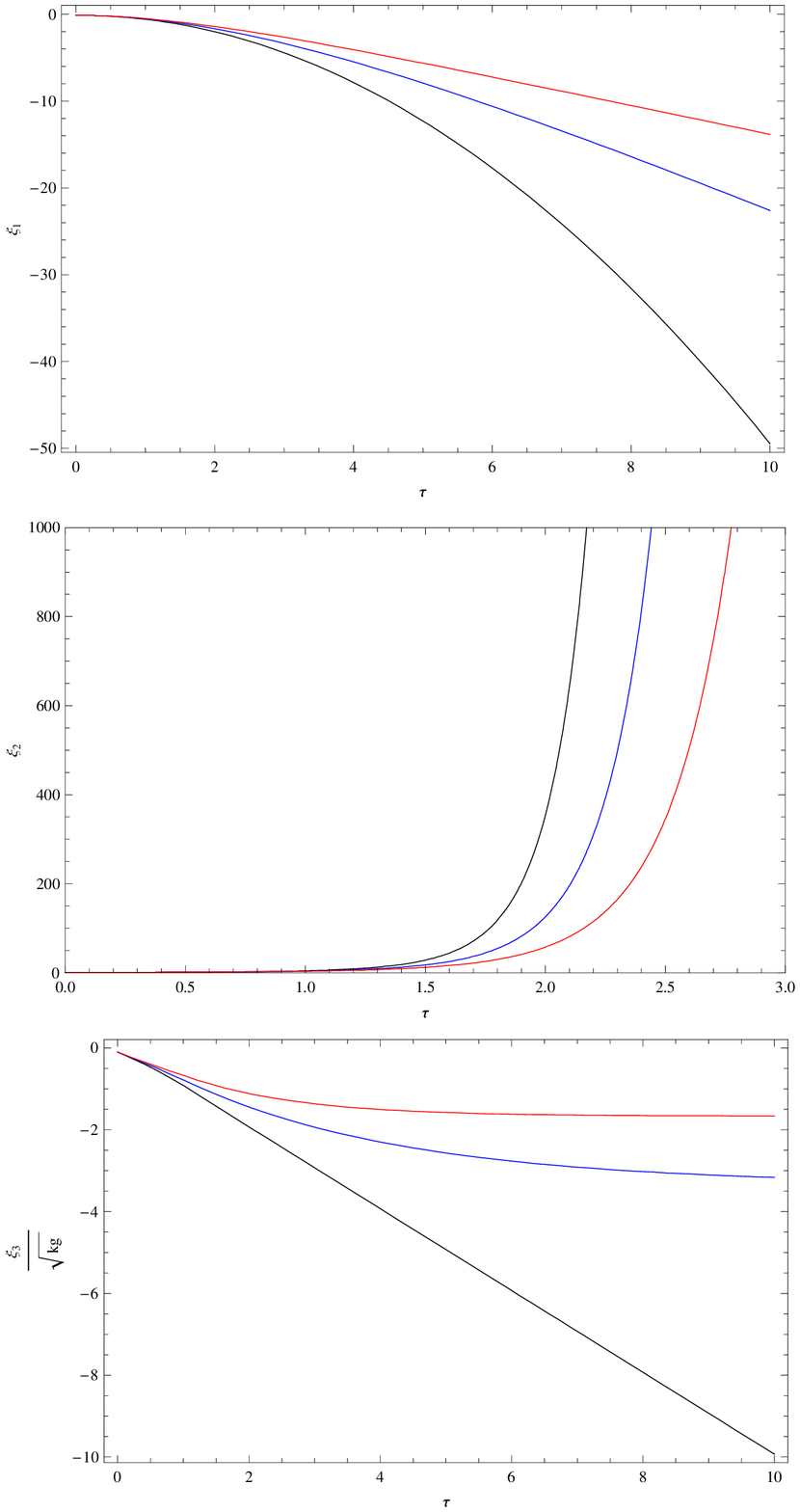}}
\begin{verse}
\vspace{-0.1cm} \caption{Variation of  $\xi_1$,$\xi_2$ and
$\xi_3/\sqrt{kg}$ with $\tau$ as obtained by the solution of the
eq.(8),eq.(9) and eq.(17) for RTI spike with initial value
$\xi_1=-0.1$, $\xi_1=1$, $\xi_3/\sqrt{kg}=-0.1$, $r=5$,$s=1/5$,
$\frac{k^2}{k_c^2}=0$ and $c_h$= 0.0 (Black), 0.1(Blue),
0.2(Red).} \label{Fig:6}
\end{verse}
\end{figure}

\begin{figure}[p]
\vbox{ \hskip 1.cm \epsfxsize=12cm \epsfbox{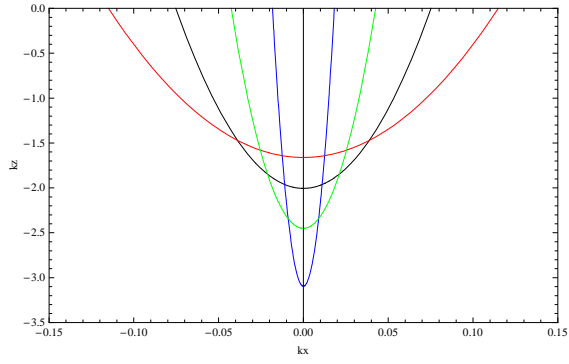}}
\begin{verse}
\vspace{-0.1cm} \caption{Formation of RTI spike where
$r=5$,$s=1/5$, $\frac{k^2}{k_c^2}=0$ and $c_h$= 0.0,
$\tau=2$(Black), $c_h$= 0.0, $\tau=2.5$(Blue), $c_h$= 0.1,
$\tau=2$(Red), $c_h$= 0.1, $\tau=2.5$(Green).} \label{Fig:7}
\end{verse}
\end{figure}
\begin{figure}[p]
\vbox{ \hskip 1.cm \epsfxsize=12cm \epsfbox{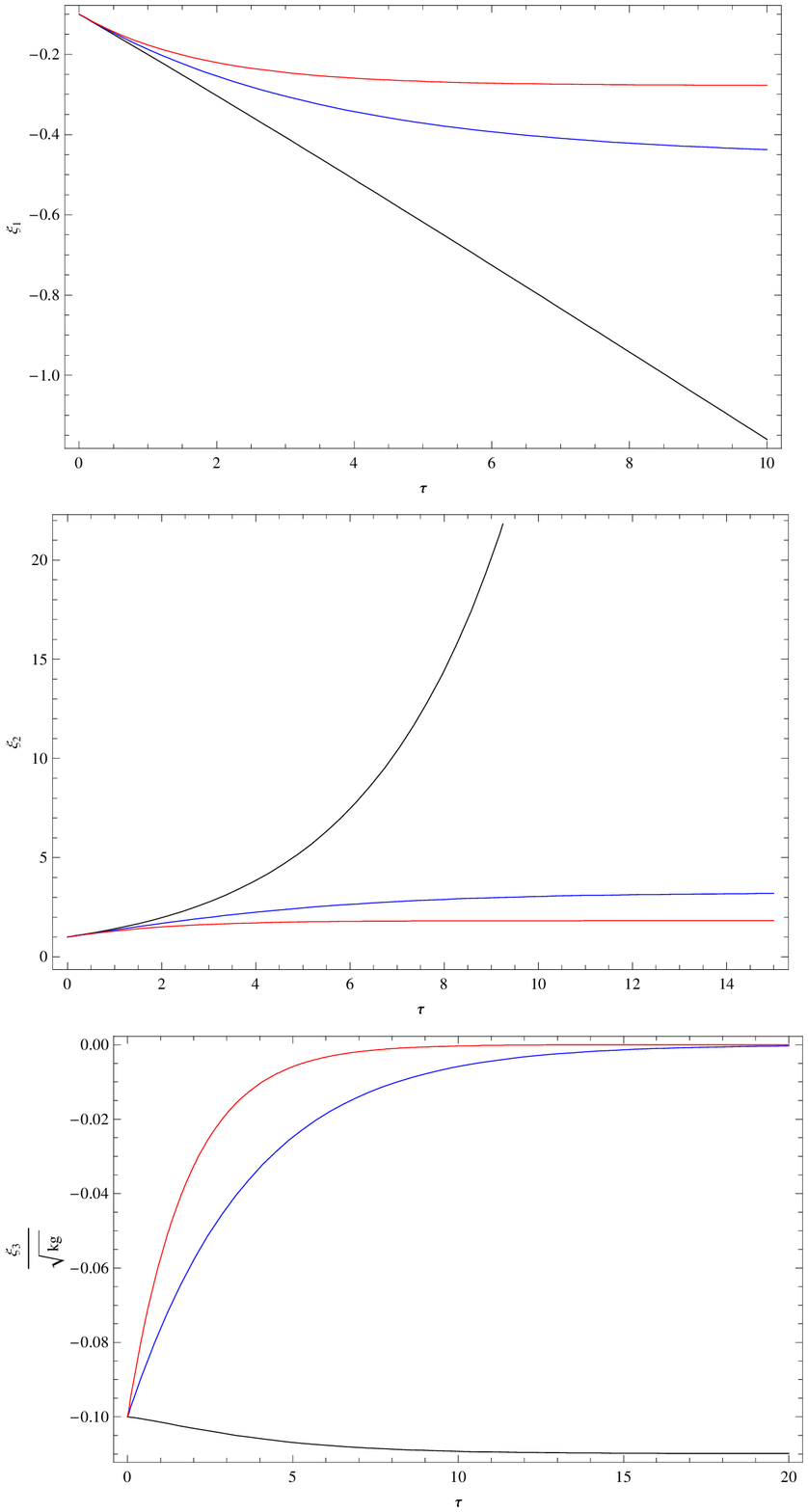}}
\begin{verse}
\vspace{-0.1cm} \caption{Variation of  $\xi_1$,$\xi_2$ and
$\xi_3/\sqrt{kg}$ with $\tau$ as obtained by the solution of the
eq.(8),eq.(9) and eq.(20) for RMI spike with initial value
$\xi_1=-0.1$, $\xi_1=1$, $\xi_3/\sqrt{kg}=-0.1$, $r=5$, $s=1/5$,
 $\frac{k^2}{k_c^2}=0$ and $c_h$= 0.0 (Black), 0.1(Blue), 0.2(Red).}
\label{Fig:8}
\end{verse}
\end{figure}

\begin{figure}[p]
\vbox{ \hskip 1.cm \epsfxsize=12cm \epsfbox{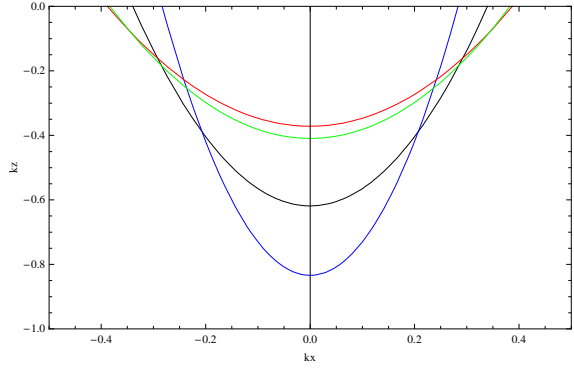}}
\begin{verse}
\vspace{-0.1cm} \caption{Formation of RMI spike where
$r=5$,$s=1/5$, $\frac{k^2}{k_c^2}=0$ and $c_h$= 0.0,
$\tau=2$(Black), $c_h$= 0.0, $\tau=2.5$(Blue), $c_h$=
0.1,$\tau=2$(Red), $c_h$= 0.1, $\tau=2.5$(Green). RTI and RMI
spike structures are different because the RTI is given by the
acceleration $g$ while RMI spike induced by shock velocity discuss
below eq.(29)} \label{Fig:9}
\end{verse}
\end{figure}
\end{document}